\documentclass[fleqn,usenatbib]{mnras}

\usepackage{newtxtext,newtxmath}
\usepackage{siunitx}
\usepackage[inline]{enumitem}
\usepackage{subfig}
\usepackage[T1]{fontenc}

\DeclareRobustCommand{\VAN}[3]{#2}
\let\VANthebibliography\thebibliography
\def\thebibliography{\DeclareRobustCommand{\VAN}[3]{##3}\VANthebibliography}

\usepackage{graphicx}	% Including figure files
\usepackage{amsmath}	% Advanced maths commands
\usepackage{amssymb}	% Extra maths symbols

\newcommand{\Rsunm}{\text{R}_{\odot}}

\title[Suprathermal Flux Peaks]{Parker Solar Probe Observations of Suprathermal Electron Flux Enhancements Originating from Coronal Hole Boundaries}

\author[A. R. Macneil et al.]{
Allan R. Macneil$^{1}$\thanks{E-mail: a.r.macneil@reading.ac.uk},
Mathew J. Owens$^{1}$,
Laura Ber{\v{c}}i{\v{c}}$^{2,3}$,
and Adam J. Finley$^{4}$
\\
$^{1}$Department of Meteorology, University of Reading, Reading, UK\\
$^{2}$LESIA, Observatoire de Paris, PSL Research University, CNRS, UPMC Universit\'e Paris 6, Universit\'e Paris-Diderot, Meudon, France\\
$^{3}$Physics and Astronomy Department, University of Florence, Sesto Fiorentino, Italy\\
$^{4}$University of Exeter, Exeter, Devon, UK
}

\date{Accepted 2020-08-28. Received 2020-08-26; in original form 2020-07-10}

\pubyear{2020}

\begin{document}
\label{firstpage}
\pagerange{\pageref{firstpage}--\pageref{lastpage}}
\maketitle
\begin{abstract}
Reconnection between pairs of solar magnetic flux elements, one open and the other a closed loop, is theorised to be a crucial process for both maintaining the structure of the corona and producing the solar wind. This `interchange reconnection' is expected to be particularly active at the open-closed boundaries of coronal holes (CHs). Previous analysis of solar wind data at \SI{1}{AU} indicated that peaks in the flux of suprathermal electrons at slow-fast stream interfaces may arise from magnetic connection to the CH boundary, rather than dynamic effects such as compression. Further, offsets between the peak and stream interface locations are suggested to be the result of interchange reconnection at the source. As a preliminary test of these suggestions, we analyse two solar wind streams observed during the first Parker Solar Probe (PSP) perihelion encounter, each associated with equatorial CH boundaries (one leading and one trailing with respect to rotation). Each stream features a peak in suprathermal electron flux, the locations and associated plasma properties of which are indicative of a solar origin, in agreement with previous suggestions from \SI{1}{AU} observations. Discrepancies between locations of the flux peaks and other features suggest these peaks may too be shifted by source region interchange reconnection. Our interpretation of each event is compatible with a global pattern of open flux transport, although random footpoint motions or other explanations remain feasible. These exploratory results highlight future opportunities for statistical studies regarding interchange reconnection and flux transport at CH boundaries with modern near-Sun missions.
\end{abstract}

\begin{keywords}
Sun: heliosphere, Sun: magnetic fields, Sun: solar wind, magnetic reconnection
\end{keywords}

\section{Introduction}\label{sec:intro}

At the time of writing, two inner heliosphere missions, \textit{Parker Solar Probe} \citep[PSP,][]{Fox2016} and \textit{Solar Orbiter} \citep[][]{Muller2013} are in operation, each having a primary goal of understanding the origins of the solar wind. Crucial to developing this understanding is the solar wind heating, acceleration, and escape from its confinement in closed magnetic fields.
Central to many models of solar wind production 
is the process of `interchange reconnection'; reconnection between one open and one closed magnetic element, which results in the opening of the previously closed loop \citep{Crooker2002}.

Interchange reconnection is a process which is necessary at the Sun in order to maintain the observed rigid rotation of coronal holes (CHs) in the face of the differential rotation of the photosphere below \citep{Nash1988,Wang1996,Wang2004,Fisk1999}. 
CHs  represent `patterns' of open magnetic flux through which, in the coronal frame, field lines are convected at the relative photospheric rotation rate \citep{Wang1996}.
The photospheric sidereal rotation rate, $\omega$, is fastest at the equator and slowest at the poles \citep[e.g.,][]{Lamb2017}. 
At high latitudes, the photosphere subrotates relative to coronal holes, so the field convects eastward in the coronal frame. At the equator, the photosphere rotates either at the same rate or slightly faster than the corona, dependent on the method used to measure the photospheric rate \citep[see][]{Bagashvili2017}, so field lines are either static or convect westward (at a lower rate than the corresponding eastward convection).
To maintain the shape of the CH boundary, footpoints of magnetic field lines which are convected into the CH must open, while those convected out must close. This opening and closing requires interchange reconnection.

It is long-established that the fast solar wind originates from CHs \citep{Krieger1973}.
Meanwhile, heavy ion charge states in the slow solar wind indicate that the plasma likely originates in closed magnetic loops \citep{Geiss1995b}.
Interchange reconnection at the CH boundary is  presented by \cite{Wang1996} as a means to release closed field plasma into the heliosphere, and contribute to the slow solar wind.
Interchange reconnection is also important in so-called `S-web' models \citep{Antiochos2007}, which posit that narrow, dynamic, open field channels, and thus open-closed boundaries where plasma can be released as above, exist throughout the corona.
The solar wind model developed by \cite{Fisk1998,Fisk1999,Fisk2003} instead introduces the concept of the diffusion of open magnetic flux through closed field regions. This  happens in order to maintain pressure balance in the corona, as a response to the motion of open flux through CHs caused by differential rotation. The open flux diffuses via repeated instances of interchange reconnection, releasing the closed field plasma to the heliosphere. Energy is released through the displacement of the open field lines.  
In this picture, the motion of open flux yields a pattern of global circulation \citep{Fisk2001}. 
Open flux is transported eastward at high latitudes by footpoint motion, and westward at low latitudes by diffusion and reconnection \citep{Fisk1999}.

Early results from PSP have brought renewed attention to interchange reconnection at the solar wind source. `Switchbacks', local reversals in the heliospheric magnetic field (HMF) accompanied by a spike in solar wind velocity (such that they are Alfv\'enic) are a striking and largely ubiquitous feature in the first PSP observations \citep{Bale2019,Kasper2019}. One candidate source of local reversals in the HMF is interchange reconnection, since newly opened field lines will be kinked \citep[e.g.,][]{Crooker2004,Owens2013b}. 
A range of other viable mechanisms to produce HMF reversals exist, including coronal jets \citep[also related to reconnection,][]{Raouafi2016,Horbury2018,Sterling2020}, stream shears \citep{Landi2005,Landi2006,Owens2018,Lockwood2019}, and solar wind turbulence \citep[e.g.,][]{Squire2020}. \textit{In situ} processes likely make a sizeable contribution to  HMF reversals observed at 0.3 to \SI{1}{AU}, since the occurrence of inverted HMF increases with solar distance over this range \citep{Macneil2020inv}. A solar origin, such as jets or interchange, remains a viable explanation for the switchbacks observed by PSP  far closer to the Sun.

\cite{Fisk2020} argue that early results from PSP's first two perihelion passes are consistent with the interchange reconnection and open flux circulation solar wind models of \cite{Fisk1998} etc. 
First, the presence and ubiquity of switchbacks are interpreted as evidence of the continuous reconnection involved in the transportation of open flux. Second, they consider the increased tangential solar wind velocity, $v_T$ (maximum $\sim\SI{50}{km.s^{-1}}$) which is far greater than predicted by current models of the solar wind \citep[e.g.,][who predict values of 1--\SI{5}{km.s^{-1}}]{Reville2020}.  This enhanced tangential flow, that was shown to be radially dependent by \cite{Kasper2019}, is recast to show that the results are also compatible with a latitudinal dependence. The low-latitude solar wind shows tangential flow consistent with the westward direction of open flux circulation at low latitudes, and this drops-off within around \SI{2}{\degree} of the heliospheric current sheet. This appears to be consistent with the development of the model by \cite{Zhao2011}, which suggested that westward transport of open flux does not extend down to the so-called `streamer-stalk' region, which underlies the heliospheric current sheet (HCS).

Signatures of interchange reconnection reported at \SI{1}{AU} may be adapted to find further evidence of this process in the PSP perihelion data. 
Here we focus on  peaks in suprathermal electron flux at stream interfaces, as reported by \cite{Crooker2010}.
Electrons in the solar wind have a suprathermal component \citep[breakpoint around 40--\SI{50}{eV} at \SI{1}{AU},][]{Bakrania2020}, which is composed primarily of a quasi-isotropic halo, and a field-aligned, anti-sunward, beam known as the 'strahl'. 
Velocity filtration arguments, in which electrons inhabiting the high energy tails of the coronal electron velocity distribution possess sufficient energy to escape into the solar wind, are one explanation for the existence of these populations in the solar wind \citep{Feldman1975,Scudder1992a,Scudder1992b,Maksimovic1997}.

Suprathermal electrons have a far longer collisional mean free path than thermal particles in the solar wind, which has lead to research on their capacity to transmit source region signatures (particularly electron temperature) to \SI{1}{AU} \citep[e.g.,][]{Hefti1999,Macneil2017}.
Evidence has emerged from PSP observations to suggest that the properties of suprathermal, and even thermal, electrons at distances $\lesssim\SI{0.3}{AU}$ are reflective of their source \citep{Halekas2020,Maksimovic2020,Bercic2020}.
In particular, core, halo, and strahl temperatures in the inner heliosphere appear to anti-correlate with solar wind speed, suggesting that at these distances, these electrons retain the electron temperature signatures of their source.

Strahl electrons at a few hundreds of eV have transit times to \SI{1}{AU} on the order of hours, as opposed to the bulk solar wind which takes $\sim2.5$--5 days.
\cite{Crooker2010} leverage this property in an analysis of the origin of peaks in pitch angle integrated suprathermal electron number flux  observed by the \textit{Wind} spacecraft \citep{Ogilvie1997} at \SI{1}{AU} during 1995, at slow to fast stream boundaries.  Comparative signatures had previously been observed in heat flux and halo temperature at interfaces by \cite{Gosling1978,Feldman1978electron}.
Slow-fast stream interfaces originate from leading CH boundaries, and are the sites of compression, as evidenced by  peaks in magnetic field,  $B$, and solar wind density, $n$ \citep[see the review by][]{Richardson2018}.
A flux peak is expected to arise at a compression as field lines are driven together and the density of the suprathermal electrons increases accordingly.  However, \cite{Crooker2010}  argue that frequent lagging or leading, by several hours, between the location of the flux peak and the peak in $B$ (their chosen signature of the interface) precludes compression as the cause of the flux peaks. It follows that the peak in flux is instead an intrinsic property of solar wind connected to the CH boundary.
Based on a prior suggestion by \cite{Borovsky2008},
the authors hypothesise that the offsets between stream interface (the convected signature of the CH boundary) and suprathermal flux peak (the argued near-instantaneous signature of the CH boundary) are a result of changing connectivity of the field to the CH boundary, through interchange reconnection. 
The tendency for offsets which are both ahead and behind the stream interface, rather than a consistent offset as expected from global circulation, may then be evidence for a `ragged' or otherwise complex open-closed boundary in the corona \citep[as in][]{Antiochos2011}. 
Alternatively, the spread in latitudes of the source CHs in their study may be responsible.

New inner-heliosphere observations by PSP present an opportunity to further investigate the source of suprathermal flux peaks at stream interfaces, and their proposed relationship with the CH boundary and interchange reconnection. 
For this purpose, near-Sun observations have several advantages over  \SI{1}{AU}. First,  stream interactions are less developed at distances $<\SI{0.3}{AU}$ \citep{Schwenn1990,Richardson2018}, so compressions are unlikely to be the cause of any peaks in flux.
With reduced compression of these regions, detailed signatures associated with the CH boundary may also be observable.
Second, if leading CH boundaries are the source of the peaks at \SI{1}{AU}, then trailing CH boundaries seem likely to also produce such peaks. The absence of such peaks at \SI{1}{AU} may be due to the fact that trailing CH boundary solar wind typically forms rarefactions, and again these are less developed near the Sun.
Finally, the studies discussed above \citep{Maksimovic2020,Bercic2020,Halekas2020} indicate that suprathermal electrons near the Sun more faithfully reflect source region properties than those at \SI{1}{AU}. Thus if the conditions of the CH boundary do intrinsically produce peaks in suprathermal flux, then examining the near-Sun solar wind provides the greatest chance of observing these peaks.
In this paper, we leverage these near-Sun advantages by analysing PSP observations of two solar wind intervals associated with CH boundaries, one trailing and one leading, in order to identify and explain the peaks in suprathermal electron flux which occur there. We use solar imagery, as well as simple solar wind mapping and coronal modelling, to understand our results in the context of the solar wind source.

\section{Data and Methods}\label{sec:datmeth}
We use publicly available PSP \textit{in situ} data\footnote{https://spdf.gsfc.nasa.gov/pub/data/psp/sweap/} from the first perihelion pass. The Solar Wind Electrons Alphas and Protons \citep[SWEAP,  ][]{Kasper2016} investigation provides all particle data for this study. Electron measurements are from the SWEAP Solar Probe Analyzers \citep[SPAN-Electrons A and B, ][]{Whittlesey2020}.
We obtain electron differential energy flux data as a function of energy, both integrated over look direction and as a function of pitch angle (PA), from the level 3 SPAN-Electrons dataset. 
We divide each of these products by the bin energy, producing the integrated differential number flux, $F$, and the pitch angle distribution of flux, $F_{\mathrm{PAD}}$. 
Figures in the next section display data from the energy 
bin at $\sim\SI{203.8}{eV}$, which is comparable to \citet{Crooker2002}.

Additionally we use the core and the strahl temperatures obtained from a fit to the level 2 SPAN-E data set. The core was modelled with a bi-Maxwellian velocity distribution function (VDF), allowing us to obtain different temperatures parallel ($T_{c\parallel}$), and perpendicular ($T_{c\perp}$) to the magnetic field direction. As the instruments’ lower-energy bins are contaminated by secondary electrons emitted from the spacecraft, the VDF data used for the core fit was limited to energy bins above  \SI{22.9}{eV}. To avoid the inclusion of the suprathermal populations an upper energy limit was set at \SI{136.6}{eV}, and the minimal pitch-angle to \SI{50}{\degree}. The strahl parallel temperature ($T_{s\parallel}$) is obtained  from a non-drifting 1D Maxwellian fit to the parallel cut through the strahl electron VDF. This fitting technique was motivated by the exospheric models' prediction that $T_{s\parallel}$, in absence of collisions and wave-particle interactions, remains unchanged from the solar corona, and preserves information about the electron VDF at its origin. For more details about the fitting procedure see Section 3.1 in \citet{Bercic2020}. 

We use level 3 solar wind ion data obtained from the Solar Probe Cup \citep[SPC, ][]{Case2020}, which includes proton velocity $\mathbf{v}_P$ in RTN coordinates, and proton number density $n_P$. 
Magnetic field data in RTN coordinates are obtained on a one minute cadence from the FIELDS \citep{Bale2016} instrument.

For context of the corona for the \textit{in situ} observations, we use full-disk solar EUV images in the \SI{193}{\angstrom} band from the\textit{ Solar Dynamics Observatory} \citep[SDO, ][]{Pesnell2015} Atmospheric Imaging Assembly \citep[AIA, ][]{Lemen2011}. We process and plot this data using the SunPy \citep{Sunpy2020} package for Python.
We use a simple two-step mapping process to link one hour averaged \textit{in situ} observations back to their `source point' at the photosphere. This method and some of its limitations are described in detail by e.g., \cite{Neugebauer1998}. Briefly, the first step assumes that the solar wind propagates with a constant, radial, velocity to map each sample from its point of observation by PSP to its origin on a Sun-centred sphere of radius \SI{2.5}{\Rsunm}.
The second step assumes that the plasma propagates entirely along the coronal magnetic field and uses a potential field source surface \citep[PFSS, see ][]{Schatten1969} model, with outer boundary set at \SI{2.5}{\Rsunm}, to link down to the source point on the photosphere along an open magnetic field line. 
The PFSS model is generated by extrapolating the magnetic field from GONG synoptic magnetograms\footnote{https://gong.nso.edu/data/magmap/archive.html} using the pfsspy package for Python \citep{pfsspy1,pfsspy2}.

\section{Results}
\subsection{Overview}
\begin{figure*}
    \centering
    \includegraphics[trim={0cm 0.1cm 0 0},clip,width = 0.75\textwidth]{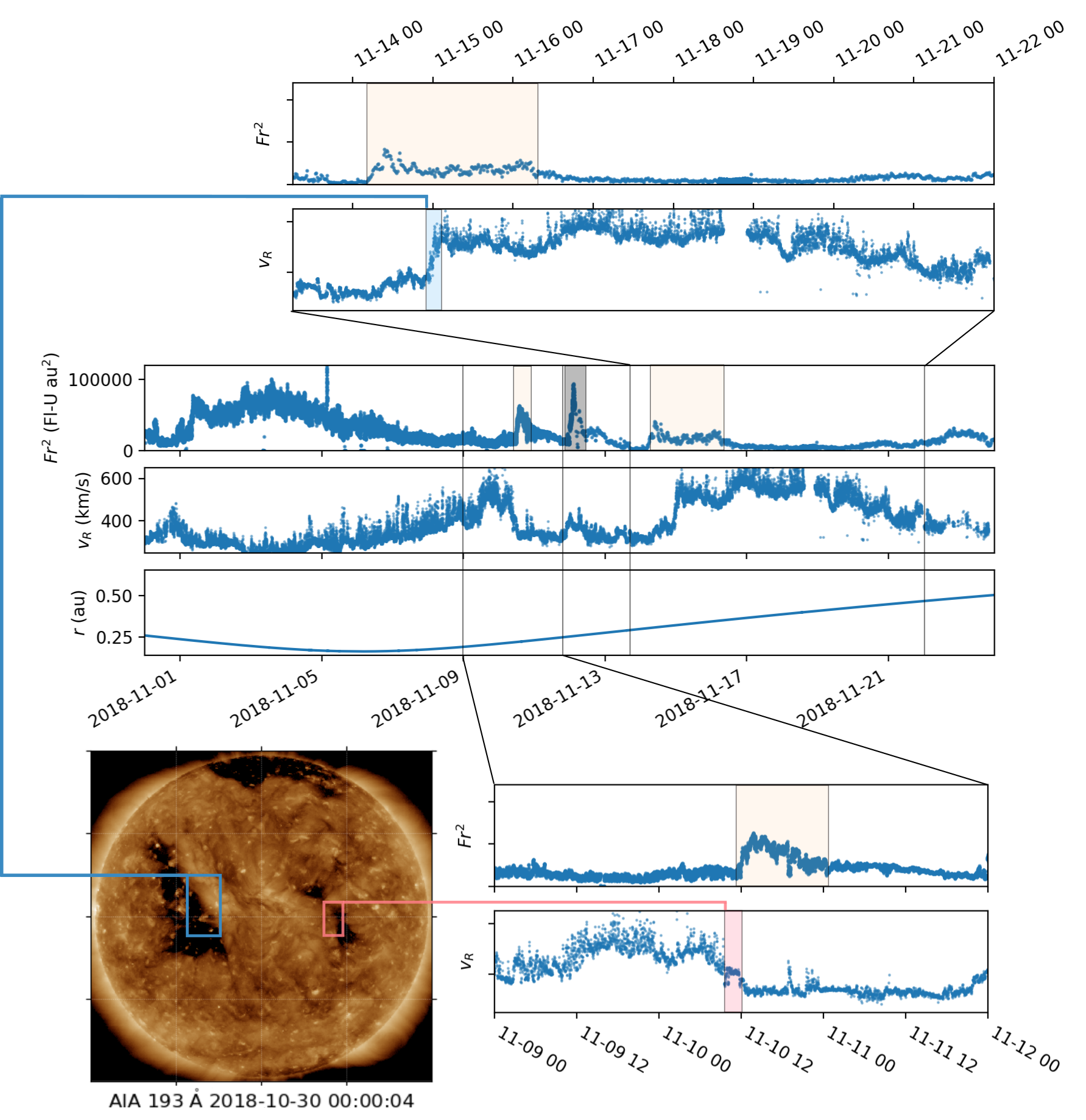}
    \caption{Summary of observations for this study. The large  time series shows pitch angle integrated  $\sim\SI{203.8}{eV}$ electron flux  (in flux units, `Fl-U', \SI{}{cm^{-2}.s^{-1}.sr^{-1}.eV^{-1}}), normalised by distance $r^2$, the radial component of the  proton velocity, and $r$. This time series covers the publicly available data for the first PSP perihelion, for solar distance $r\lesssim\SI{0.5}{AU}$, 
    The two peaks in electron flux which are the subject of this study are highlighted in orange. A third peak is greyed-out, which corresponds to an ICME (see text). The two periods of study are each magnified above and below the large time series.
    A rough estimate of the stream interface (slow-fast for top, fast-slow for bottom) are highlighted blue for the top interval and red for the bottom. 
    The bottom left corner of the figure shows an SDO-AIA image of the predicted CH sources for the two \textit{in situ} periods. The image time corresponds to roughly half a solar rotation prior to the estimated time of solar wind release for these periods.
    Boxed regions highlight the CH boundaries which we expect to be the source of the stream interfaces for the respective \textit{in situ} intervals they are connected to.
    }
    \label{fig:summary}
\end{figure*}

This study focuses on a pair of solar wind intervals encountered by PSP during the outward-moving phase of its first perihelion encounter; after the period of super-corotation. Figure \ref{fig:summary} summarises the first perihelion, the two intervals, and their predicted solar sources. The \textit{in situ} data shown are integrated $\sim\SI{203.8}{eV}$ electron flux, normalised by solar distance $r^2$, and proton radial velocity $v_R$.
This time series is shown for the full span of publicly available data for the first perihelion, for $r\lesssim\SI{0.5}{AU}$. We highlight three strong localised peaks in flux, each of which occur following perihelion. The peak which is highlighted grey occurs during a small interplanetary coronal mass ejection (ICME), so we discount it from this study. The two peaks highlighted in orange are associated with the largest rapid changes in $v_R$ in this interval, making them candidates for velocity transitions resulting from CH boundaries. These peaks and their surrounding regions are the two intervals which we select for study. Other flux enhancements occur during this perihelion, but these are either relatively small (e.g., the two peaks at the beginning of the interval) or are not associated with a clear velocity transition (e.g., the long-lived enhancement which takes place during the extended low-speed interval before perihelion, or the final peak shown in the chosen time period).

Of the selected intervals, one contains a slow stream followed by a fast stream (slow-fast transition, top of figure) and the other a fast stream followed by a slow stream (fast-slow transition, bottom of figure). 
The slow-fast and fast-slow wind transitions are likely due to leading and trailing boundaries, respectively, of different coronal holes. 
In addition to the above features, these intervals are well suited to this study because
\begin{enumerate*} 
\item the intervals exemplify both  a leading and trailing CH boundary,
\item each interval maps to a CH/CH boundary (see below), and
\item a brief period of slow wind, and at least one day of fast wind, precedes/follows the relevant transition, allowing for comparisons of CH boundary properties with `pure' wind on either side.
\end{enumerate*}

The EUV image from SDO-AIA in the centre of Figure \ref{fig:summary} shows a pair of equatorial coronal holes which are predicted by the two-step mapping procedure to be the source regions for the two selected stream boundaries. (Detailed mapping results are shown in Section \ref{sec:map}.) 
Note that this imagery corresponds to when these source CHs were Earth-facing; around half a solar rotation prior to the release of the PSP-observed plasma.
The small, western, CH is the reported source of the solar wind observed by PSP during the innermost period of perihelion 1, which has been extensively studied \citep[e.g.,][]{Badman2020,Bale2019,Kasper2019,Allen2020}.
The trailing (leading) boundary of the eastern (western) CH is the likely source of the fast-slow (slow-fast) transition. Magnetogram data from SDO-HMI (not shown) shows that the two coronal holes are of opposite polarity, and fall on opposite sides of the HCS.

We highlight the slow-fast velocity interface in the top panel in  blue. This interface has been studied in detail and confirmed to be a stream interaction region (SIR) by \cite{Allen2020}. An enhancement of energetic particles is found at this stream interface by \cite{Cohen2020}, who note that it is likely a result of particle acceleration at greater solar distances.
The precise location of the region corresponding to the leading CH boundary is not as well-defined near the Sun as at \SI{1}{AU}, where the CH boundary is typically part of the stream interface itself \citep{Schwenn1990,Borovsky2016} and is therefore sharp and clear in time. 
Furthermore, while a CH boundary near the Sun can be defined as the location of open-closed flux cutoff, \cite{McComas2002} argue for a CH boundary `layer' of finite thickness, across which coronal plasma properties shift continuously. The proximity to some theoretically sharp open-closed magnetic boundary from which solar wind plasma must have originated to be considered as CH boundary plasma is not clear. 
This is particularly true for any suprathermal flux peak signatures which are associated with the boundary (or proximity to it).
For our example, we highlight the slow-fast transition at the leading stream interface in blue, but suggest that plasma either side of it can be reasonably argued to be associated with the CH boundary.

The stream interface in a fast-slow transition is difficult to identify at \SI{1}{AU} \citep{Borovsky2016} but should be clearer close to the Sun, before rarefactions grow large.  Based on suggestions that CH boundary plasma is intermediate between fast and slow speeds \citep{McComas2002},  we highlight the brief period where the velocity stagnates between the clearer fast and slow periods as the velocity transition, and the most probable location of CH boundary plasma. This is similar to the inflection point in velocity which \cite{Borovsky2016} found to be an indicator of the trailing stream interface in \SI{1}{AU} observations.
As above, solar wind surrounding the highlighted region could also reasonably be associated with a CH boundary source.

The strong peaks in integrated suprathermal flux are present close to the highlighted slow-fast and fast-slow transitions, so are likely associated with the CH boundaries. The peak at the slow-fast transition spans either side of the velocity increase, lasting for around \SI{2.5}{days}. Meanwhile the peak at the fast-slow transition appears to follow the velocity transition, when the wind speed is around \SI{300}{km.s^{-1}}, and persists for around \SI{12}{hours} as it gradually tails-off.

\subsection{\textit{In Situ} Analysis}\label{sec:insitu}

\begin{figure*}
    \centering
    \includegraphics[clip,trim={0 .5cm 0 0},width=0.75\textwidth]{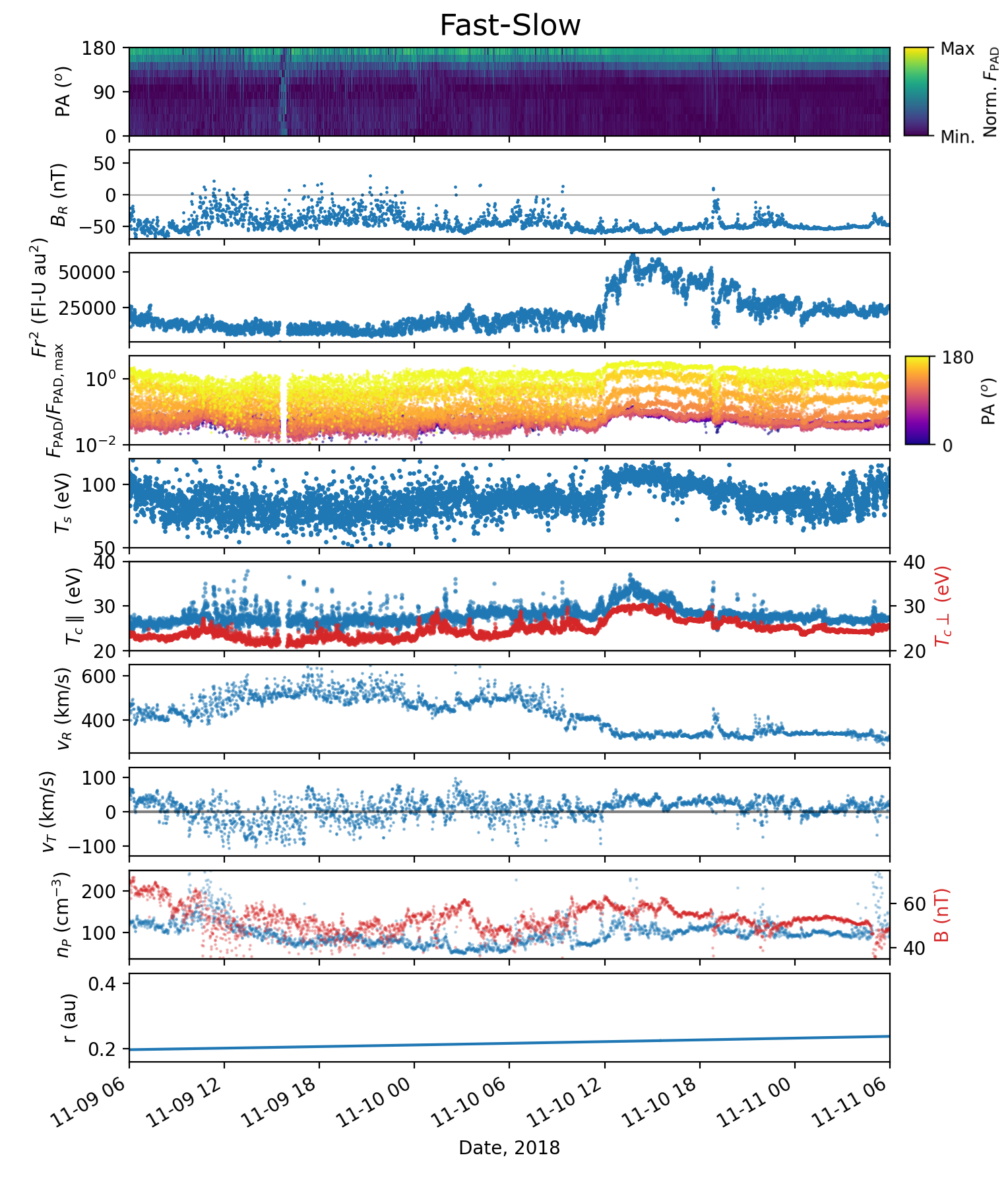}
    \caption{Detailed \textit{in situ} PSP time series of the fast-slow (this page) and slow-fast (next page) intervals, cropped to focus on the stream interfaces and flux peaks. From top to bottom the panels show: column-normalised heatmap of $\log{F_{\mathrm{PAD}}}$ at $\sim\SI{203.8}{eV}$; minutely radial magnetic field component $B_R$; $Fr^2$ in flux units; the pitch angle distribution components normalised by the maximum value in each interval, $F_{\mathrm{PAD}}/(F_{\mathrm{PAD, max}})$, at \SI{203.8}{eV} for 12 PA bins (indicated in colour bar); strahl temperature $T_{s\parallel}$; core parallel (perpendicular) temperature $T_{c\parallel}$ ($T_{c\perp}$); proton radial velocity $v_R$;  proton tangential velocity $v_T$;  proton density $n_p$ and minutely magnetic field magnitude $B$; and solar distance $r$. Start and end times of the enhancements in suprathermal electron flux are indicated by dashed lines.
    }
    \label{fig:insitu}
\end{figure*}

\begin{figure*}\ContinuedFloat
    \includegraphics[clip,trim={0 .5cm 0 0},width=.75\textwidth]{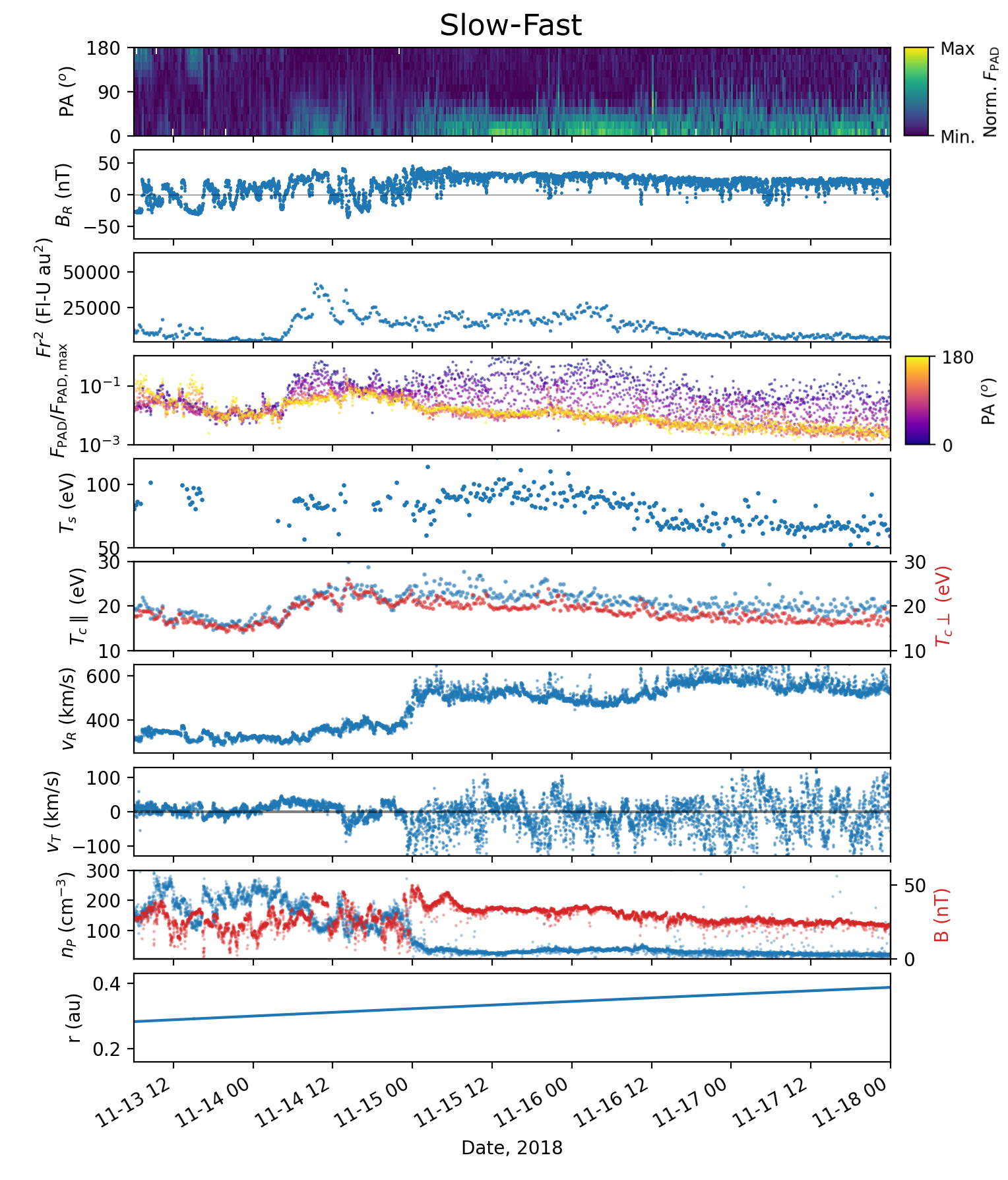}
    \caption{(continued).}
\end{figure*}

Figure \ref{fig:insitu} shows the \textit{in situ} properties of the two periods of interest in detailed time series. Here we summarise the key results.
The flux peak in the fast-slow interval occurs across all pitch angle bins (shown by $F_{\mathrm{PAD}}/F_{\mathrm{PAD, max}}$) and so involves both halo and strahl. The peak during the slow-fast interval is more complex. Halo and strahl peak together prior to the velocity transition. Then afterwards, the flux in the halo bins drops off, but the strahl enhancement remains, and drives the peak in integrated flux. 
These profiles for the flux peaks are consistent in bins of energy up to around \SI{1}{keV} (not shown). Below energies of around \SI{85}{eV}, evidence of peaks remain, but the overall profile begins to depart from that seen at \SI{203.8}{eV}. Particularly at the slow-fast transition, low-energy flux appears to track with density $n_p$ (see below). 
Based on the duration of the peaks and PSP orbital data, the size of the region of peaked halo and strahl flux in the fast-slow interval is around \SI{2}{\degree} in longitude. The corresponding region in the slow-fast interval,  where halo and strahl peak together prior to the velocity transition is larger: \SI{7}{\degree} in longitude.

We include core and strahl temperatures to support the flux results. $T_{s\parallel}$ for the fast-slow interval peaks similarly to the \SI{203.8}{eV} flux. In the slow-fast interval, there are missing $T_{s\parallel}$ data points due to strahl drop outs during the flux peak, as shown in the PADs. Near the time where the integrated flux peak drops off (second dashed line) $T_{s\parallel}$ also decreases.
Parallel and perpendicular core temperatures, $T_{c\parallel}$ and $T_{c\perp}$, both peak at the fast-slow transition. These peaks ramp up more slowly than $T_{s\parallel}$, and also sharply cease several hours before the gradual drop in the suprathermal flux.
At the slow-fast transition, $T_{c\parallel}$ and $T_{c\perp}$  increase at the point where the suprathermal flux peak begins, but then drop off more gradually.

The fast streams in each interval fall on either side of the HCS and thus have different polarity (as evidenced by $B_r$ and electron PADs in the top two panels of Figure \ref{fig:insitu}). HMF reversals are present during the flux peaks in both the fast-slow and slow-fast intervals. These are not as frequent as the switchbacks found in their respective preceding/following fast streams. The large reversal in particular during the slow-fast interval is notable due to the dropout of strahl which accompanies it.
 
Proton density, $n_p$, and magnetic field magnitude, $B$, are both structured and variable during the peaks in flux. However, there are no large increases in $n_p$ or $B$  which coincide well with the timing of the flux peaks, and so could explain them in terms of simple compression.
During both periods where halo and strahl flux peak together, proton tangential velocity $v_T$ shows large positive values. Mean $v_T$, $\langle v_T \rangle = \SI{24.5}{km.s^{-1}}$ when calculated over the fast-slow flux peak, while $\langle v_T \rangle= \SI{0.76}{km.s^{-1}}$ for the preceding fast stream, and $\langle v_T \rangle = \SI{12}{km.s^{-1}}$ for the following slow stream. Similarly, $\langle v_T \rangle = \SI{8.3}{km.s^{-1}}$ for the flux peak preceding the slow-fast transition, contrasting with  $\langle v_T \rangle =\SI{-12.4}{km.s^{-1}}$ in the fast stream following it and $\langle v_T \rangle =\SI{4.6}{km.s^{-1}}$ in the slow stream preceding it.
During this period of enhanced $+v_T$ there is a sizeable $-v_T$ excursion, which coincides with the HMF reversal and strahl dropout.

\subsection{Detailed Mapping and PFSS Results}\label{sec:map}
%This is made using  /Code/Connection/bmcarr.py
\begin{figure}
    \centering
    \includegraphics[clip,trim={5.3cm .3cm 5.3cm 1.cm},width=.49\textwidth]{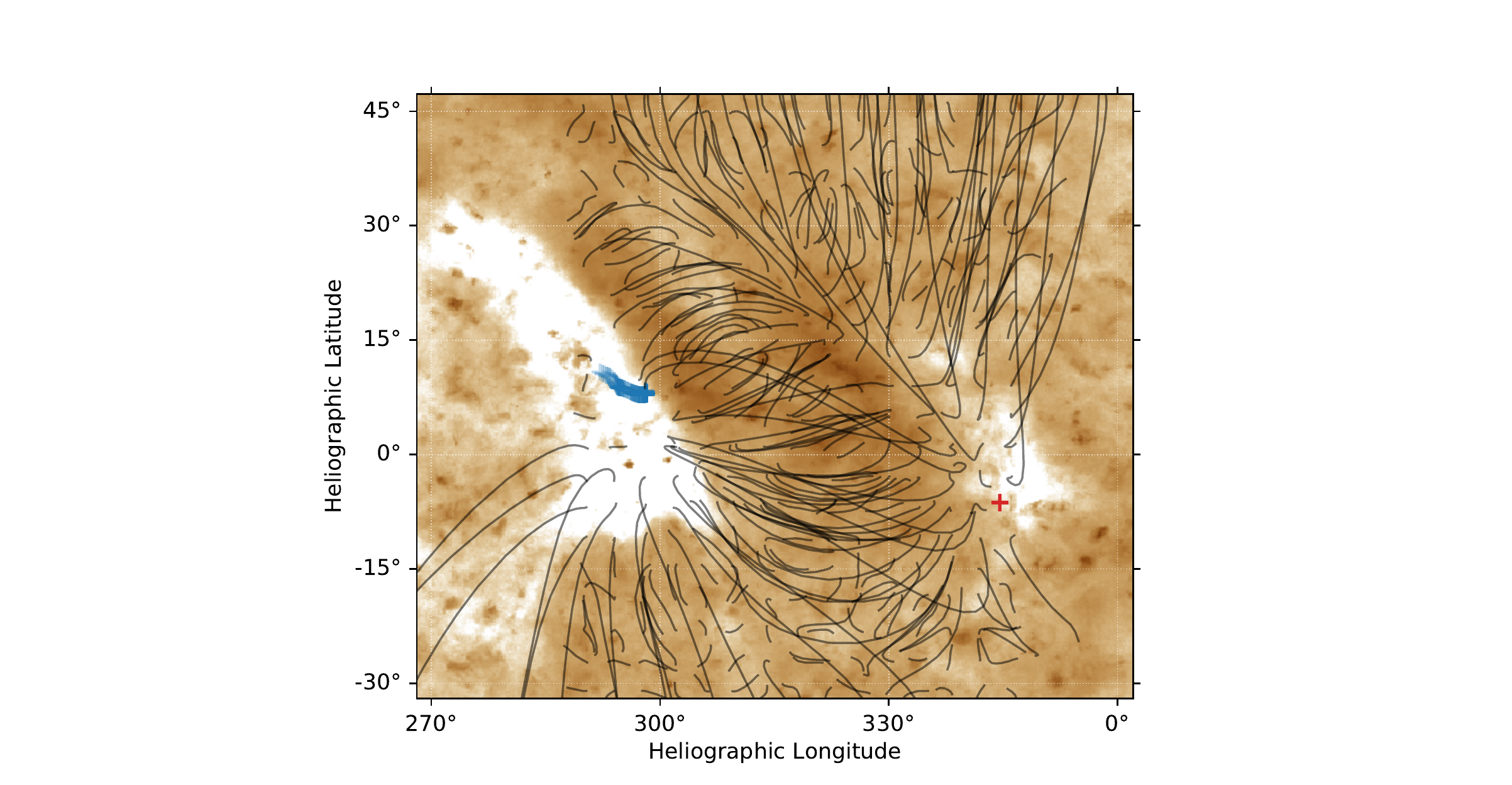}
    \caption{Sub-region of a Carrington map constructed from AIA-\SI{193}{\angstrom} images over the time range 2018-10-25 to 2018-11-19 as described in the text, in Carrington heliographic coordinates. The sub-region of the image shown is constructed primarily from three images from 2018-10-27T16:00:04, 2018-10-30T00:00:04, and 2018-11-01T06:00:04. The colours have been reversed compared to those in Figure \ref{fig:summary}.
    Overlaid on the map are field lines and PSP mapped source points, derived from a PFSS model generated from the GONG synoptic magnetogram for 2018-11-13T00:14. 
    Red source points correspond to solar wind mapped from the earlier fast-slow interval, while red points are those mapped from the later slow-fast interval.
    Black lines represent the closed magnetic field lines which have at least one foot point rooted in the box with corners $[\mathrm{lon, lat}] = [\SI{290}{\degree},\SI{-45}{\degree}], [\SI{345}{\degree},\SI{45}{\degree}]$. Red crosses represent the mapped source points of solar wind observed by PSP during the fast-slow and slow-fast intervals in Figure \ref{fig:insitu}.}
    \label{fig:detailmap}
\end{figure}

Figure \ref{fig:detailmap} provides more detailed observations and modelling of the source region for the two streams of interest. 
The figure shows a coronal EUV map, projected into Carrington heliographic coordinates, of the equatorial coronal holes highlighted in Figure \ref{fig:summary}. This is a sub-map of a full Carrington rotation map, which we generate by reprojecting 12 AIA-\SI{193}{\angstrom} images which are evenly spaced in time over the course of the period 2018-10-25 to 2018-11-19. The reprojected images are combined by taking mean pixel values, where pixels which were measured close to disk centre are preferentially weighted over those measured close to the limb.
The image times for this sub-map were chosen to show the source CHs as observed in the rotation prior to the PSP observations. SDO observations of these CHs for the rotation following PSP observations (e.g., on 2018-11-26) show that a small active region (NOAA AR-12728) has emerged off the trailing boundary of the eastern CH. STEREO A EUVI-\SI{195}{\angstrom} observation of these locations at around 2018-11-18 also precede the emergence of this active region, and occur after the \textit{in situ} observations in this study. 
Our chosen image dates are thus representative of the corona during the period of solar wind release, which is prior to the emergence of the AR.

Crosses overplotted on Figure \ref{fig:detailmap} show the source  locations produced by following the mapping procedure described in Section \ref{sec:datmeth}  for the two periods of interest shown in Figure \ref{fig:insitu}.  The PFSS model used for the mapping is generated using the GONG magnetogram made closest to the middle of the time range over which the PSP-encountered solar wind was released (as predicted by the initial step of the mapping).
Blue source points are mapped from the slow-fast interval, and fall within the large eastern CH, starting at the leading edge near \SI{0}{\degree} latitude and `moving' north-east. Meanwhile, all red source points,  mapped from the slow-fast interval, are associated with the trailing boundary of the smaller western CH. These represent several days of data, although they appear as only a single point at the current resolution.
This mapping is broadly consistent with that obtained using similar techniques by \cite{Badman2020}. %check 
While the red source points appear to fall on the brighter boundary of the source CH, rather than the CH itself, this may be a result of misalignment between the modelled open flux locations from the PFSS model, and the CH boundary appearance in these \SI{193}{\angstrom} images. Such disagreement could be due to the difference between our choice of time for the GONG magnetogram and AIA images. 
The key result is that solar wind plasma encountered by PSP, likely during the fast streams in the two periods of interest, originates in these CHs. The slow-fast transition thus corresponds to the eastern CH's leading boundary, and the fast-slow transition to the western CH's trailing boundary, supporting the association between these regions highlighted in Figure \ref{fig:summary}.

We compute magnetic field lines derived from the PFSS model on a \SI{4}{\degree} resolution grid on the photosphere. Black lines overlaid on Figure \ref{fig:detailmap} represent the subset of these field lines which are both closed, and have at least one footpoint fall within the box of corners $[\mathrm{lon, lat}] = [\SI{290}{\degree},\SI{-45}{\degree}], [\SI{345}{\degree},\SI{45}{\degree}]$, which we choose to probe the closed magnetic field between the two CHs. In this region, large loops are approximately east-west oriented, and cross the polarity inversion line (as verified by the GONG magnetogram) making up the streamer belt. These east-west loops are rooted all along the leading (trailing) boundary of the eastern (western) CH, including at locations near the mapped solar wind source.

\section{Discussion}
We have analysed time series and source regions for a pair of suprathermal electron flux peaks which are associated with solar wind velocity transitions. These two peaks are prominent features in the overall evolution of suprathermal flux during the first PSP encounter. 
The most obvious \textit{in situ} cause of suprathermal flux peaks in general is compression regions, as discussed by \cite{Crooker2010} and recounted here in Section \ref{sec:intro}. Our results indicate that compression is not the cause of either of our example flux peaks, because there are no associated enhancements in magnetic field or density.  The flux peaks are likely instead due to observed concurrent suprathermal temperature enhancement, as \cite{Crooker2010} argued for the peaks at \SI{1}{AU}.
Further, these peaks both occur close to the Sun, where SIRs are typically underdeveloped in comparison to \SI{1}{AU}.
The  peak at the fast-slow transition in particular also takes place at the typical location for solar wind rarefaction, rather than compression. 
Finally, it is also unlikely that the peaks are due to reflection of electrons from e.g., a downstream shock, since both are strongly driven by the (outward travelling) strahl and not localised to the stream interfaces.
The explanation that these peaks in flux are instead a result of some feature and/or process in the corona is thus likely.

As verified by both solar imagery and mapping (Figures \ref{fig:summary} and \ref{fig:detailmap}), the fast streams  during each interval originate in two different CHs, and so the velocity transitions in each interval are a result of their respective CH boundary.  While enhancement in electron heat flux and temperature have been observed to be a function of solar wind speed in the PSP data \citep[e.g.,][]{Halekas2020}
we do not believe these peaks to be a result of this phenomenon. This is because periods of comparable solar wind speed (and other parameters) immediately follow the flux peak in the fast-slow interval, and immediately precede the peak in the fast-slow interval, but do not show enhancement in suprathermal flux, or the related core and strahl temperatures.
We therefore conclude that the most likely explanation of these peaks is due to their association with the CH boundaries themselves.
While other enhancements in flux are observed during the PSP encounter (Figure \ref{fig:summary}), this is expected based on previous studies relating suprathermal electron properties (such as temperature) to different sources \citep{Hefti1998,Macneil2017,Bercic2020,Maksimovic2020}. Thus CH boundaries are not the sole cause of enhancements in suprathermal electron flux close to the Sun, and this is largely unsurprising.

The above conclusions broadly match those drawn by \cite{Crooker2010}, based on their statistical study of peaks associated with slow-fast transitions. 
Here, the additional association of core temperature peaks with the suprathermal ones may be a result of thermal electrons near the Sun being closely related to their source region properties (Section \ref{sec:intro}).
While we do not have the same statistical basis as \cite{Crooker2010}, the presence of these peaks in the absence of strong compression, and at a fast-slow transition in particular, is a novel result which can only be obtained with near-Sun data from PSP.

\subsection{Extension and Shifting of Suprathermal Peaks}

\begin{figure*}
    \centering
    \includegraphics[width=0.8\textwidth]{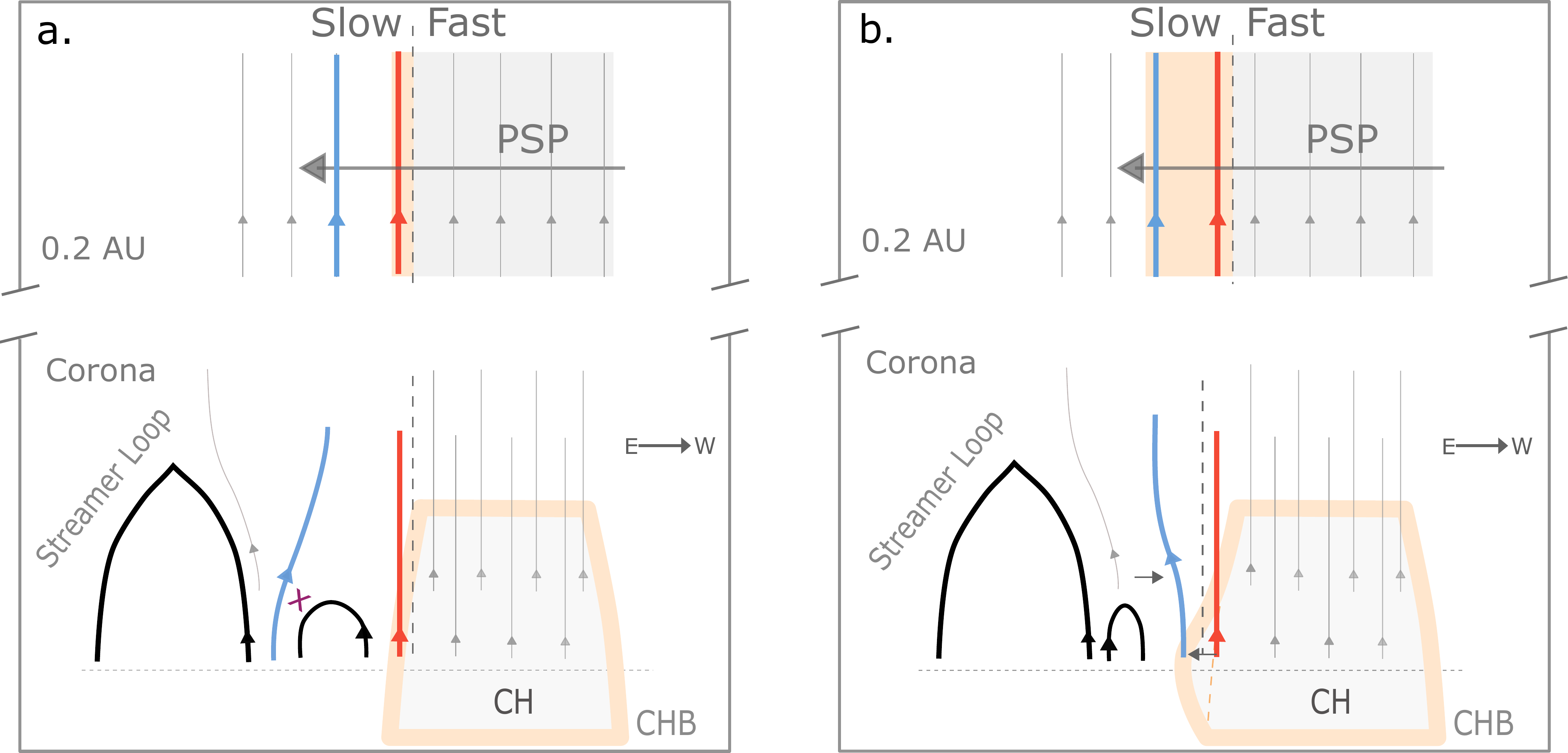}
    \caption{Schematic of a possible CH boundary reconnection process in the corona, and corresponding effect on solar wind observed by PSP near \SI{0.3}{AU}. 
    Panel\textbf{ a}:  Pre-reconnection configuration in corona. An eastern (trailing) CH boundary (CHB) lies adjacent to the streamer belt and closed loops there. We highlight one field line in  in red which is rooted in the region at the CHB which is assumed to produce enhanced electron flux (shown in orange). 
    We also highlight an open field line in blue which is embedded somewhere in the nominally closed flux region.
    A dashed vertical line shows the assumed boundary between fast and slow wind production. 
    PSP's path through the solar wind produced from the coronal configuration, a fast-slow transition, is shown above. The stream interface is located at the dashed line. The same highlighted field lines are shown. The orange background highlights the region connected to the CHB, where enhanced suprathermal flux is observed \textit{in situ}.
    Panel\textbf{ b:} Post-reconnection configuration in corona. The blue highlighted field line reconnects with an adjacent small loop, transferring open flux west towards the CH. Flux adjacent to the CHB opens as a result, effectively shifting the boundary eastward. The blue field line at PSP is now rooted in the CHB. PSP's path is shown shortly after this reconnection takes place. The peak in flux is shifted back/extends into the fast stream (see text) because the blue highlighted field line is now rooted in the enhanced suprathermal flux region, and suprathermal electrons rapidly propagate outwards. The location of the velocity transition, and any other signatures which do not propagate as quickly as the suprathermal electrons, remain in the same location as in Panel a.
    }
    \label{fig:schem}
\end{figure*}

In Section \ref{sec:insitu} we found that the halo and strahl flux, and core and strahl temperature, all peak  on the slow side of each velocity transitions. 
These peaks thus appear to either originate directly from the CH boundary, or from just outside of it, depending on how the velocity transition itself relates to the boundary. Some mismatching features can be found in the peak profiles for the different parameters in each interval.
For the fast-slow interval, the peaks in $T_{c\parallel (\perp)}$ fall off sharply, relative to those in the suprathermal parameters which persist longer by several hours and decay gradually. 
For the slow-fast interval, the integrated suprathermal flux peak spans either side of the velocity transition, but the halo and strahl contribute differently on each side.
Each of these apparently mismatching features  may result from changes at the source region which is communicated rapidly by the suprathermal electrons \citep{Borovsky2008,Crooker2010} as discussed in Section \ref{sec:intro}.

Figure \ref{fig:schem}, based on Figure 5 of \cite{Crooker2010}, is a schematic illustration of one way in which a change at the source could produce the above mismatching features at the fast-slow transition.
A detailed description of the schematic is contained in the figure caption.
In this example, interchange reconnection at the CH boundary, in which open flux is transferred westward away from the streamer belt, effectively  shifts the location of the boundary to the east. As a result, enhanced suprathermal electron fluxes are released onto fields which PSP encounters following the original peak location. Signatures convected more slowly to the spacecraft, such as the peaks in core temperature, and the velocity transition, still occur at the original location. 
The suprathermal peaks thus appear shifted, or extended, relative to the core temperature peaks, which corresponds to PSP's observations.

The presence of the suprathermal peaks on either side of the slow-fast transition could also indicate a change of connectivity (if the `original' peak location was on the slow side of the velocity transition, as we observe for the fast-slow interval). Westward transfer of flux by interchange reconnection, as shown in Figure \ref{fig:schem} but now at the leading CH boundary, could lead to the peak extending to the fast side of the transition. This peak manifesting in strahl, but not the halo, may be a result of the strahl's more rapid propagation to PSP.
The peaks in $T_{c\parallel (\perp)}$ are not necessarily consistent with this, however, since the $T_{c\parallel (\perp)}$ enhancement gradually falls off across the velocity transition, whereas we might expect it to end sharply before it. One explanation for this could be that the greater solar distance of this interval means that core electron properties are not so well correlated with the source as in the fast-slow interval.   
An alternative explanation, which does not involve interchange reconnection, is that bulk plasma both before and after the slow-fast velocity transition originates from the CH boundary, or at least from the location where some process produces the electron peaks. This is feasible since, as discussed earlier, the precise location of CH boundary plasma is not well-established this close to the Sun.

The westward transfer of open flux, which we have argued could explain the features of the flux peaks at each transition,  is consistent with the direction of open flux circulation at equatorial latitudes predicted by \cite{Fisk1999,Fisk2001}. However, we cannot confirm a systematic effect with just these two examples. 
The continual reconnection at CH boundaries is likely more complex than we have represented schematically, so many configurations which produce this signature likely exist.
The inferred peak shifts could equally be due to reconnection with randomly oriented closed loops \citep{Fisk2001} or at a ragged boundary  \citep[][]{Antiochos2011}. \cite{Crooker2010} made a similar suggestion when they found the peaks in flux at \SI{1}{AU} to not exhibit a systematic offset from the SIR. Understanding the true cause of these features will require the comparison of several more CH boundary streams at similar distances, which will be possible later in the PSP mission.

\subsection{Origins of Peaks in Suprathermal Flux}

We can speculate on the possible mechanisms causing the peak in suprathermal electron flux originating from the CH boundaries.
Under exospheric models, an increase in flux may result from increased electron temperature at the source. 
Enhanced source region electron temperature is a likely cause of the flux peaks we observe, since they are concurrent with \textit{in situ} $T_{s\parallel}$ and $T_{c\parallel (\perp)}$ peaks.
The CH boundary is expected to have  electron temperature intermediate between slow wind source regions and the CH proper \citep[e.g., based on charge state measurements,][]{McComas2002}. The enhanced flux at the boundary relative to the fast wind would then be an expected result of this temperature relation. However, the fact that we observe the suprathermal flux to be greater in the CH boundary than adjacent slow wind periods is not consistent with this explanation. 

Non-thermal processes, such as magnetic reconnection, have previously been invoked to explain the energisation  of electrons in the corona \citep[e.g.,][]{Che2014a,Yang2015}.
Since CH boundaries are favourable locations for continuous interchange reconnection,  it may be involved in explaining the enhanced suprathermal flux. The enhancement could, for example, result from the release of hot material from newly-opened loops, although again this may not explain why the CH boundary flux is greater than the slow wind flux. Alternatively, energy released from the reconnection process itself may allow for the heating and subsequent escape of electrons, producing the enhanced flux. Energy deposition to the plasma is an expected result of the interchange solar wind models \citep{Zhao2011}. This explanation is highly dependent on sufficient energy being released through this reconnection, and the mechanisms through which it may be transferred to the electrons being viable in these locations.

Some evidence of source region interchange reconnection  can be found in the relevant \textit{in situ} intervals (aside from the shift in peak location discussed above).
For example, magnetic reversals are a possible reconnection signature which are present during both peaks in flux. However, they are also present outside of the flux peak intervals, during the fast streams in each period. 
Additionally, the tendency for strong $+v_T$ at the CH boundaries/flux peaks could result from the westward transport of flux due to diffusion, enabled by reconnection \citep{Fisk2001}.
While deflections at stream boundaries are common at \SI{1}{AU} and beyond, these arise from stream interactions \citep[e.g.,][]{Crooker2012b,Borovsky2016} which will be less developed so close to the Sun. The typical profile of $v_T$ at these interactions is to increase in magnitude prior to the stream interface, change sign rapidly at it, and then trail off. The $v_T$ enhanced intervals do not exhibit this profile. Instead the transport of flux around CH boundaries likely provides a non-radial component to the solar wind flow, which is a local property that has implications for evaluating the angular momentum-loss rate of the Sun (Finley \textit{et al.} 2020, in prep.). This transport would appear to be necessarily stochastic within CHs, since we do not observe the strongly enhanced $v_T$ throughout the fast solar wind streams.  
These two CH boundaries which exhibit enhanced $v_T$ lie on either side of the HCS. $v_T$ being greater in these regions than in the neighbouring slow wind, from nearer the HCS, could be a result of the absence of open flux transport in the streamer-stalk region, as suggested by \cite{Zhao2011}.

A final notable feature is the HMF reversal(s) during the flux peak preceding the slow-fast velocity transition. One reversal in particular is coincident with a strahl dropout, and a strong negative $v_T$ excursion in the midst of the otherwise positive interval. 
Negative $v_T$ events are of particular interest, as they could be a mechanism by which the Sun regulates the release of angular momentum in response to the enhancements in angular momentum caused by the previously discussed footpoint motion.
In this case, the HMF reversal may result from an interchange event in the corona, and the strahl drop out caused by scattering or reflection of electrons on the kinked structure between the source and spacecraft.
Alternatively, the drop out could indicate a disconnection event \citep[e.g.,][]{Gosling2005b} caused by pinch-off reconnection upstream of the spacecraft, or some other topological change. The field reversal would then be due to newly-reconnected field convecting over PSP, and the $-v_T$ would indicate the propagation of this Alfv\'enic disturbance away from the reconnection site. The size of the disturbance in $-v_T$ is indeed on the order of the local Alfv\'en speed.  However, many other features of the clear disconnection event identified by \cite{Gosling2005b}  (oppositely oriented strahl, event centred on the HCS crossing) are not present here.
Related to this, \cite{Owens2011} predicted an increase in disconnection events in the case where there is a strongly inclined HCS, as we observe here in  Figure \ref{fig:detailmap}.

We note that at the time of writing, evidence of possible discrepancies between $v_T$ data as measured by the SPC and SPAN instruments (the latter of which were not available during the first perihelion pass) is emerging with more recent PSP encounters. 
Until the source of this discrepancy is understood, the enhancements in $v_T$ reported here are thus subject to some uncertainty, as is their potential role in the processes discussed above. 
The primary results and conclusions of this study do not hinge upon the $v_T$ observations, and so should not be subject to such uncertainty.

\section{Conclusions}

We have presented two instances of temperature-driven peaks in suprathermal flux, observed by PSP, which occur in a pair of solar wind streams originating from two different coronal hole boundaries. 
Arguments based on plasma properties (particularly electron moments), the observation of a peak at a trailing CH boundary (when historically peaks were reported at leading boundaries) and the close solar distance of PSP (reducing transport effects) suggest that these peaks are not the result of solar wind dynamics but instead are intrinsic to the coronal hole boundary source, as first suggested by \cite{Crooker2010}. 
In such a case, the relative positions of the different suprathermal peaks  could be a result of changing connectivity to the dynamic CH boundary.
While the mechanism which might produce these peaks in electrons from the coronal hole boundary is not clear, the involvement of reconnection, which is understood to be highly active there, seems likely.

The interpretation of these observations is closely linked to the motion of open magnetic flux on the Sun 
and the origins of the solar wind. Determination of whether the global circulation of flux,  random motion, or some other effect, is driving the apparent discrepancy between different peaks in suprathermal flux and electron temperatures will require careful identification of any systematic offset. 
Confirmation of all of the above conclusions and suggestions thus requires statistical evidence.      
This can be acquired in the future through analysis of a diverse spread of near-Sun leading and trailing CH boundary streams in the growing PSP dataset.

%do I have funding for Laura to add?
\section*{Acknowledgements}
Work was part-funded by Science and Technology Facilities Coun-cil (STFC) grant No. ST/R000921/1, and Natural Environment Re-search Council (NERC) grant No. NE/P016928/1.
AJF acknowledges funding from the European Research Council (ERC) under the European Union’s Horizon 2020 research and innovation programme (grant agreement No 682393 AWESoMeStars).
We acknowledge the NASA Parker Solar Probe Mission and SWEAP team led by Justin Kasper for use of data. We also acknowledge the FIELDS team lead by Stuart Bale for use of data.
This work utilizes data from the National Solar Observatory Integrated Synoptic Program, which is operated by the Association of Universities for Research in Astronomy, under a cooperative agreement with the National Science Foundation and with additional financial support from the National Oceanic and Atmospheric Administration, the National Aeronautics and Space Administration, and the United States Air Force. The GONG network of instruments is hosted by the Big Bear Solar Observatory, High Altitude Observatory, Learmonth Solar Observatory, Udaipur Solar Observatory, Instituto de Astrof\'{\i}sica de Canarias, and Cerro Tololo Interamerican Observatory.
This research made use of Astropy,\footnote{http://www.astropy.org} a community-developed core Python package for Astronomy \citep{astropy:2013,astropy2018}. 
This research used version 1.1.1 \citep{Mumford2020} of the SunPy open source software package \cite{sunpy_community2020}.
Figures 1--3 were produced using the Matplotlib plotting library for Python \citep{Hunter2007}.

\section*{Data Availability Statement}
The data underlying this article are publicly available at the following locations. PSP SWEAP data: \url{https://spdf.gsfc.nasa.gov/pub/data/psp/sweap/}, PSP FIELDS data: \url{https://spdf.gsfc.nasa.gov/pub/data/psp/fields/}, GONG synoptic magnetograms:  \url{https://gong.nso.edu/data/magmap/archive.html}, SDO AIA images and HMI magnetograms: \url{https://sdo.gsfc.nasa.gov/data/}.

\bibliographystyle{mnras}
\bibliography{PSP} % if your bibtex file is called example.bib

\appendix

% Don't change these lines
\bsp	% typesetting comment
\label{lastpage}
\end{document}